%% file: main.tex
\documentclass[review]{elsarticle}

\usepackage{graphicx}
\usepackage{hyperref}
\usepackage{bm}
\usepackage{wasysym}
\usepackage{float}
\usepackage{lineno}
\modulolinenumbers[5]
\usepackage{makecell}
\input{commands.tex}

\journal{Journal of \LaTeX\ Templates}
\begin{document}
\title{\bf {A forward-angle large-acceptance magnetic spectrometer}}
\author[jlab]{B.~Wojtsekhowski} \corref{*} \ead{bogdanw@jlab.org}
\author[uva]{G.~Cates}
\author[infn]{E.~Cisbani}
\author[jlab]{M.~Jones}
\author[cmu]{G.~Franklin}
\author[uva]{N.~Liyanage}
\author[jlab]{L.~Pentchev}
\author[uconn]{A.J.R.~Puckett}
\author[jlab]{R.~Wines}
\address[jlab]{Thomas Jefferson National Accelerator Facility, Newport News, VA 23606}
\address[uva]{University of Virginia, Charlottesville, VA 23606}
\address[infn]{INFN Sezione di Roma and Istituto Superiore di Sanit\`a, Rome, Italy}
\address[cmu]{Carnegue Mellon University, Pittsburg, PA 23606}
\address[uconn]{University of Connecticut, Storrs, CT 06269}

\cortext[*]{Corresponding Author. Tel: +1 757 269 7191}
\date{\today}
\include{abstract}      %%%%%%include%%%%%%%
\maketitle
%
%\tableofcontents
\large
%\linenumbers


\include{introduction}	% overview of magnetic spectrometers in the field
% of electron scattering and the most prominent examples of them
%
\include{mechanical}		% formulation of spectrometer parameters: solid angle, scattering angle, field intergral

\include{magnet}		% magnet, coils, field characteristics in gap and on beamline

\include{parameters}	% optical parameters, solid angle vs angle, momentum acceptance, angular range
\include{detectors}		% open detector configuration, GEM and calorimeter 
% rate capability, calorimeter concept; proton and neutron polarimeter
%
\section*{\large Summary}

The SBS magnetic spectrometer with solid angle up to 70~msr capable of operating at high luminosity of a few 10$^{38}$ cm$^{-2}$/s and central angle of the spectrometer down to 15~degrees was constructed at the Thomas Jefferson National Accelerator Facility.
A main part of the spectrometer is a dipole magnet with a horizontal magnetic field and a narrow horizontal cut in the yoke. 
The cut made in one side of the yoke close to the pole allows us to obtain beamline to dipole mid-plane angles 
as small as 15 degrees without loss of the spectrometer solid angle.
The remaining field on the beamline is reduced by two-layer magnetic shielding with an external layer made of iron rings.
A number of physics experiments have already used this spectrometer, and a
significant physics program of future experiments has already been approved.

\section*{\large Acknowledgments}
We appreciate the support given to this project by K.~de~Jager, L.~Cardman, R.~McKeown, R.~Ent, and C.~Keppel. 
The contributions of A.~Gavalya, W.~Seay, and Ch.~Soova to the spectrometer design and J.~LeRose to studies of the magnetic optics are very much appreciated.
We are thankful to Y.~Zhilichev for helpful instructions on the OPERA magnetic field calculation package.

This material is based upon work supported by the U.S. Department of Energy, Office of Science, Office of Nuclear Physics under contracts DE-AC05-06OR23177 (TJNAF), DE-FG02-01ER41168 (G.C.), DE-FG02-03ER41240  (N.L.), DE-FG02-87ER40315  (G.F.), and DE-SC0021200  (A.P.), also supported in part by the EU Horizon 2020 Research and Innovation Program under the Marie Sklodowska-Curie Grant Agreement No. 101003460 (E.C.).

\bibliography{main.bib}

\end{document}

%% file: commands.tex
\usepackage{scalerel}
\usepackage{tikz}
\usetikzlibrary{svg.path}

\definecolor{orcidlogocol}{HTML}{A6CE39}
\tikzset{
  orcidlogo/.pic={
    \fill[orcidlogocol] svg{M256,128c0,70.7-57.3,128-128,128C57.3,256,0,198.7,0,128C0,57.3,57.3,0,128,0C198.7,0,256,57.3,256,128z};
    \fill[white] svg{M86.3,186.2H70.9V79.1h15.4v48.4V186.2z}
                 svg{M108.9,79.1h41.6c39.6,0,57,28.3,57,53.6c0,27.5-21.5,53.6-56.8,53.6h-41.8V79.1z M124.3,172.4h24.5c34.9,0,42.9-26.5,42.9-39.7c0-21.5-13.7-39.7-43.7-39.7h-23.7V172.4z}
                 svg{M88.7,56.8c0,5.5-4.5,10.1-10.1,10.1c-5.6,0-10.1-4.6-10.1-10.1c0-5.6,4.5-10.1,10.1-10.1C84.2,46.7,88.7,51.3,88.7,56.8z};
  }
}

\newcommand\orcidicon[1]{\href{https://orcid.org/#1}{\mbox{\scalerel*{
\begin{tikzpicture}[yscale=-1,transform shape]
\pic{orcidlogo};
\end{tikzpicture}
}{|}}}}

\usepackage{hyperref}

\usepackage{graphicx}% Include figure files
\usepackage{dcolumn}% Align table columns on decimal point
\usepackage{bm}% bold math
\usepackage{subfigure}
\usepackage{epstopdf}
\usepackage{hyperref}
\usepackage{color}
\usepackage[utf8]{inputenc}

\newcommand{\gep}{GEp}

\newcommand{\gevsq}{(GeV/$c$)$^2$}
\newcommand{\qsq}{Q$^2$}

\newcommand{\Bp}{$B_\perp$}
\newcommand{\BpI}{$\int{B_\perp}$}

%% file: abstract.tex
\begin{abstract}

A large solid angle magnetic spectrometer for high luminosity and forward scattering angles was constructed at the Thomas Jefferson National Accelerator Facility.
A number of physics experiments have used this spectrometer, and a significant physics program of future experiments has already been approved.
A key feature of the spectrometer concept is a horizontal slit opening that allows the beamline to pass through the yoke of the spectrometer magnet.
This design enables a short distance between the target and spectrometer, resulting in a 70~msr solid angle acceptance.
The residual magnetic-field on the beamline inside the slit is reduced by a two-layer magnetic shielding system, with the external layer comprising a set of iron rings.
Two correcting magnets, before and after the dipole, were used to compensate for the transverse component of the fringe field outside of the dipole yoke.
The mechanical stability of the tall dipole magnet in close proximity to the target was provided by means of a heavy counterweight. 
\end{abstract}

%% file: introduction.tex
\section{Introduction}
\label{sec:introduction}

Magnetic spectrometers have been developed for many experiments in electro- photo-nuclear physics~\cite{Hofstadter,Taylor,DESY,Enge,Bertozzi,NIKHEF,Alcorn,SHMS,Blomqvist}.
Their critical characteristics include: momentum and angular resolution, a range of settings achievable for the central trajectory's momentum and angle, a range of momentum and solid angle acceptance about the central trajectory for a given setting, and the operational luminosity.
The solid angle of a typical universal spectrometer of 5-7~msr has been achieved for a several-GeV momentum range~\cite{Alcorn, SHMS} and up to 28~msr for lower momenta~\cite{Blomqvist}.
The relative momentum resolution varies from 10$^{-3}$ to $\sim 3\times10^{-4}$ in the spectrometers referenced above with momentum acceptance of 10-20\%.
The typical angular resolution is about 1~mrad. 

These spectrometers have a significant bend angle of the particle trajectories of 20-45 degrees and a well shielded detector packages allowing operation at very high electron-nucleon luminosities up to 10$^{39}$~cm$^{-2}$/s.
The minimum scattering angle for these spectrometers is typically larger than 10~degrees due to the significant width of the spectrometer quadrupole and dipole magnets. 
An additional septum magnet for horizontal deflection of the trajectories has been used for reaching scattering angles down to 5$^\circ$ with a modest solid angle of 3-5~msr~\cite{SHMS,sHRS}.
The specialized low-resolution high-momentum spectrometers~\cite{Petratos} had 50\% momentum acceptance and operated at an angle of 4.5-7$^\circ$ with a solid angle up to 0.6~msr.

For experiments such as pion electro-production, which often have a lower momentum resolution requirement, dipole-only non-focusing spectrometers were used, see e.g. Refs.~\cite{Bonn, Gross, Doug}.
Those spectrometers provided a much larger solid angle up to 50-100~msr, a wide momentum range starting from a few hundred MeV/c, and a relative momentum resolution of (0.3-1)$\times$p[GeV]\%.
The acceptable luminosity was up to 10$^{37}$~cm$^{-2}$/s.

Investigation of processes like $(e,e^\prime h)$ at high momentum transfer \qsq~and large $z \,=\, p_h/p_{\gamma^*}$ (where $h$ is for a hadron and $\gamma^*$ is a virtual photon) is of high interest for hadron physics~\cite{Barabanov,Achenbach}.
A productive experiment at high \qsq~requires maximizing the product of luminosity and detector acceptance. 
The device whose magnetic/mechanical design is described in this paper, the Super Bigbite Spectrometer, SBS, was originally proposed for measurement of the ratio of the electric to magnetic form factors of the proton at high momentum transfer~\cite{GEp,GEp-update}.

The SBS spectrometer addresses the need for a device with large solid angle and modest momentum resolution adequate for the exclusive (and semi-exclusive) reactions with a nucleon.
It allows large momentum acceptance and small central angle with respect to the incident beam.

The measurement of form factor ratios in the proton electric form factor experiment, GEp, is based on the polarization transfer method \cite{Akhiezer,Scofield,Recalo,Arnold}, which allows us to mitigate complications related to the small contribution of \gep~to the elastic scattering cross section at high momentum transfer, but requires a very large statistics due to the low analyzing power of the proton polarimeter at high proton momenta~\cite{Basiliev,Azhgirey}.
In addition to the GEp experiment, the SBS spectrometer has become a basic part of several other experiments~\cite{GMn, GEn, TDIS, nTPE, GEn-RP, pWACS, ALL, GEp+} and fits well into an investigation of the semi inclusive deep inelastic scattering process~\cite{SIDIS}. 
A small scattering angle and large solid angle in the hadron arm are essential for a wide physics program~\cite{BW2014}.

The momentum reconstruction in the SBS uses the so-called "vertical bend" approach, in which the magnet midplane is vertical.
The vertical coordinate of the beam-target interaction point is defined by the location of the beam, so the track's coordinates and directions after the magnet permit unique reconstruction of the particle momentum, $p$, the polar and azimuthal angles of the particle trajectory at the target relative to the beam direction, $\theta$ and $\phi$,  and the $y$ coordinate of the scattering vertex (transverse to the dispersive direction) at the target, see e.g. Ref.~\cite{Optics}.

A detector in magnetic spectrometers is protected from low energy charged particles produced in the target by a strong magnetic field.
However, photon background is also present.
The detector package used in those spectrometers is usually based on multi-wire chambers, scintillator counters with good time resolution, and Cherenkov counters and shower calorimeters for particle identification.
At higher luminosity operation of the tracking detectors can be complicated because of loss of the track reconstruction efficiency due to high occupancy in the tracking detector produced by the photons.

The SBS detector package includes several layers Gas-Electron-Multiplier chambers (GEM)~\cite{Sauli,GEM}, which have gain stability for rate well above 10~MHz/cm$^2$, and a large hadron calorimeter~\cite{HCAL}.

This paper is organized as follows: 
After the introduction~\ref{sec:introduction}, section~\ref{sec:mechanical} presents the spectrometer mechanical design.
Section~\ref{sec:Magnet} describes the magnetic elements of the spectrometer including the beamline shielding and the fringe field dipole correctors.
Section~\ref{sec:parameters} provides the spectrometer parameters.
The following section~\ref{sec:Detectors} describes the layout of the detectors in the spectrometer for the GEp experiment.

%% file: mechanical.tex
\section{\large Mechanical structure}
\label{sec:mechanical}

The spectrometer, see Fig.~\ref{fig:3D-view}, is based on:
\begin{figure}[!htb]
\includegraphics[trim = 5mm 30mm 150mm 30mm, angle=0, width=0.5\textwidth]{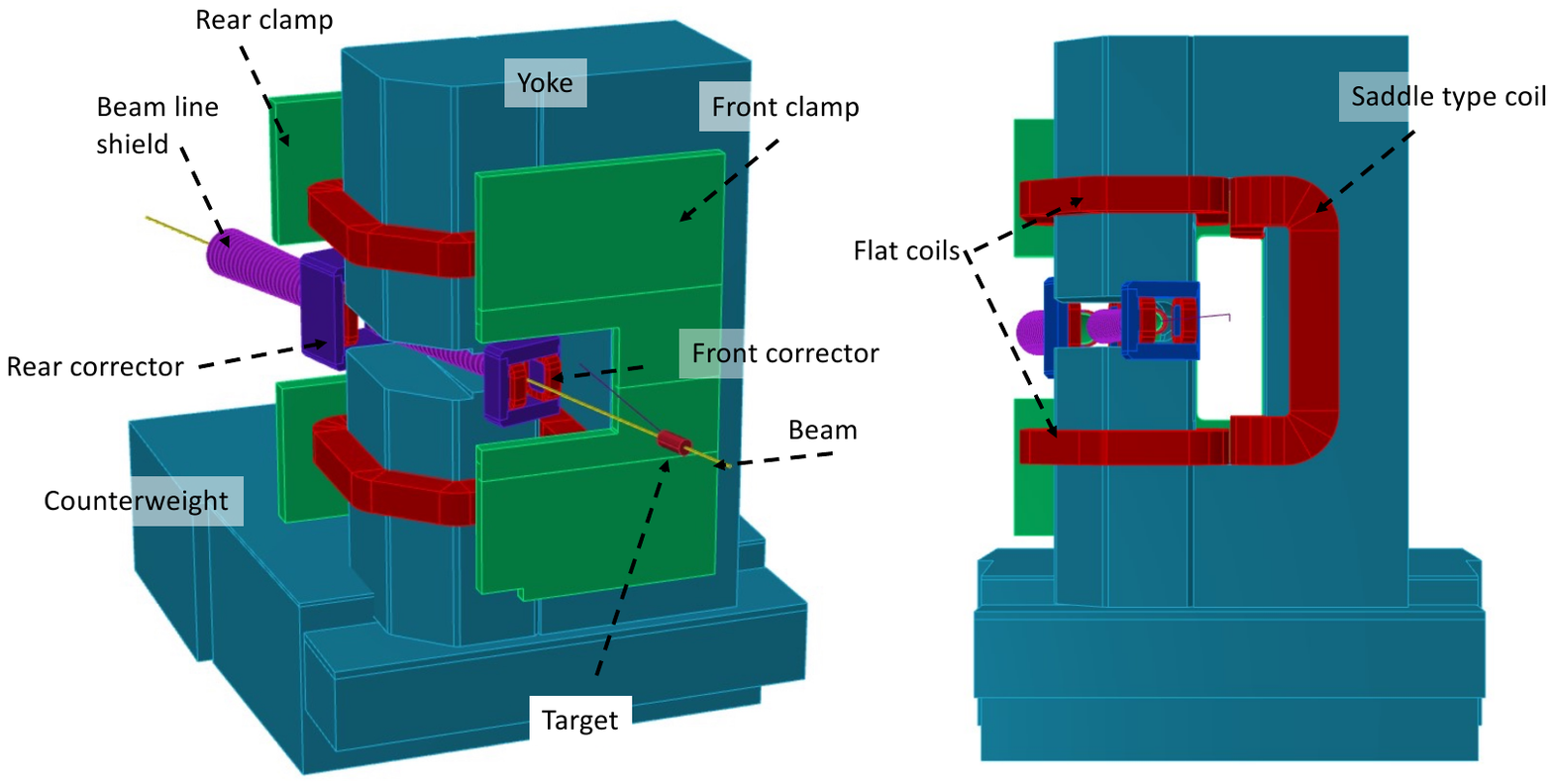}
\caption{The views of the SBS magnet. A 3D view on the left and front view on the right (a front field clamp removed for visibility of structure).}
\label{fig:3D-view}
\end{figure}
\begin{itemize}
\item A large aperture dipole magnet with iron-dominated horizontally oriented magnetic field.
\item A slit shape opening in the magnet yoke for the beam to pass.
\item A double layer magnetic shield of the beamline with a multi-ring-based outer shield. 
\end{itemize}

\begin{figure}[!htb]
\includegraphics[trim = 30mm 50mm 20mm 70mm, angle=0, width=1.\textwidth]{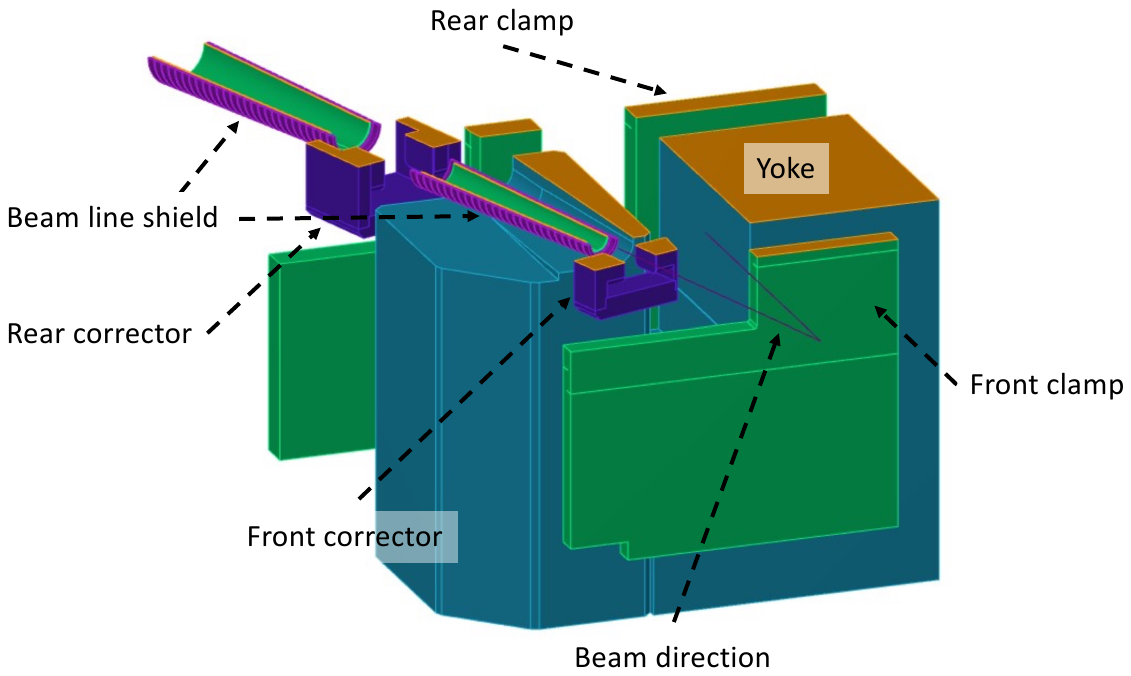}
\caption{The SBS magnet with removed upper portion for visibility of the beamline. Outer shielding of the beamline (rings) shown in purple for visibility.}
\label{fig:SBS-cut}
\end{figure}

The design allows us to create an opening for the beamline, as shown in Figs~\ref{fig:3D-view} and \ref{fig:SBS-cut}.
As a result, we can place the magnet at a short distance
from the target even at a small angle between the beam and spectrometer (160~cm between the target and yoke front face). 

The configuration of the dipole with a cut in the yoke has a common feature with the Lambertson magnet~\cite{Lambertson}, which is well known in the field of accelerator design where the low-field region on the beam orbit of the accelerator should be separated from the area with a strong field of the septum magnet used for extraction/injection of the beam.

The development of a high coordinate-resolution GEM-based tracking detector~\cite{GEM} allows us to use a relatively small field integral of 2~Tesla-meters for momentum analysis of particles with momentum up to 8~GeV/c.
The vertical bend allows us to locate the whole tracking detector behind the magnet at a distance of 4-6~m from the target and reduce dramatically the rate induced by the low-energy charged particles.

The vertical bend dipole configuration of the spectrometer, in contrast to a solenoidal option, allows good momentum resolution for small scattering angles with the tracking detector located behind the magnet. 

The component of the magnetic field along the beamline does not present a big concern in an experiment, but the transverse one does;
it can deflect the beam out of the beam dump.
The transverse field also deflects the low energy electrons produced in the target, which leads to an increase in the background rate in the detectors.
In many cases the transverse field on the beamline can be reduced sufficiently by enclosing the outgoing beam with an iron pipe.
However, in the case of SBS, there is also a longitudinal field which leads to saturation of the pipe shield, so shielding of the transverse field
is also lost.
In a double-layer pipe type shielding, both layers are also saturated. 
We found a solution in the form of a ring-based configuration of the outer layer of the shield, see Fig.~\ref{fig:BeamShield} below.
\begin{figure}[!htb]
\includegraphics[trim = 40mm 40mm 130mm 70mm, angle=0, width=0.6\textwidth]{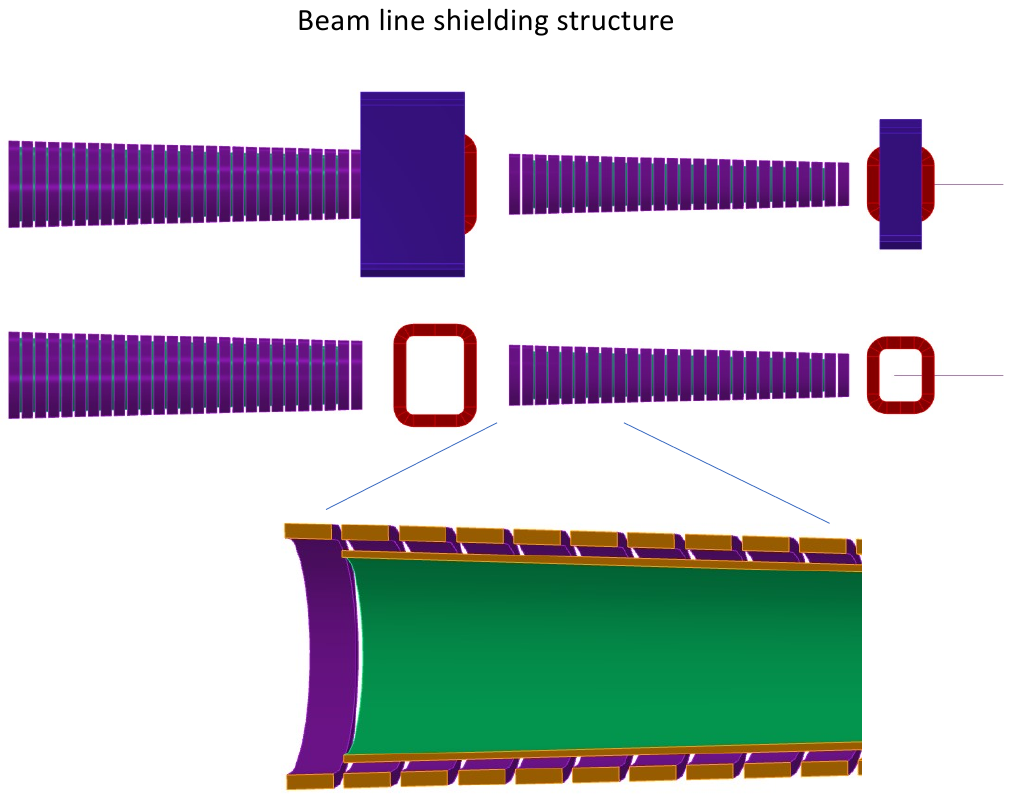}
\caption{The side view of the beamline magnetic shield.
In the middle picture, the iron of the correctors was removed.
The upstream portion of the beam shield is shown in the bottom picture (with a bit of rotation) and zoomed with the closer half of the shield removed for visibility of the structure.}
\label{fig:BeamShield}
\end{figure}

Mechanical stability of tall magnets traditionally was accomplished by the use of a wide support structure.
However, such a structure puts restrictions on the magnet location and leads to loss of solid angle. 
We realized a different approach with dipole supported in a cantilever configuration with a heavy counterweight.

As in many dipole magnets we have used coils on both sides of the magnet gap.
On the side far from the beamline we used a saddle type coil, see the right picture in Fig.~\ref{fig:3D-view}.
We used two flat coils on the side close to the beamline because the experiments require a wide range of spectrometer angles relative to the beamline.
This unavoidably leads to a significant fringe field on the opposite side of the beamline where the second arm detector is usually located.

The 100 ton dipole yoke for this spectrometer and its counterweight were constructed from 
%BNL's 
"48D48" magnets which were donated by Brookhaven National Laboratory.
%BNL.

%% file: magnet.tex
\section{\large Magnetic aspects of the spectrometer}
\label{sec:Magnet}

The magnetic field analysis/design of the SBS was performed using the TOSCA/OPERA package~\cite{OPERA}. 

\subsection{\large Magnetic field of the dipole}

\begin{figure}[!htb]
\includegraphics[trim = 50mm 35mm -10mm 25mm, angle=0, width=1.5\textwidth]{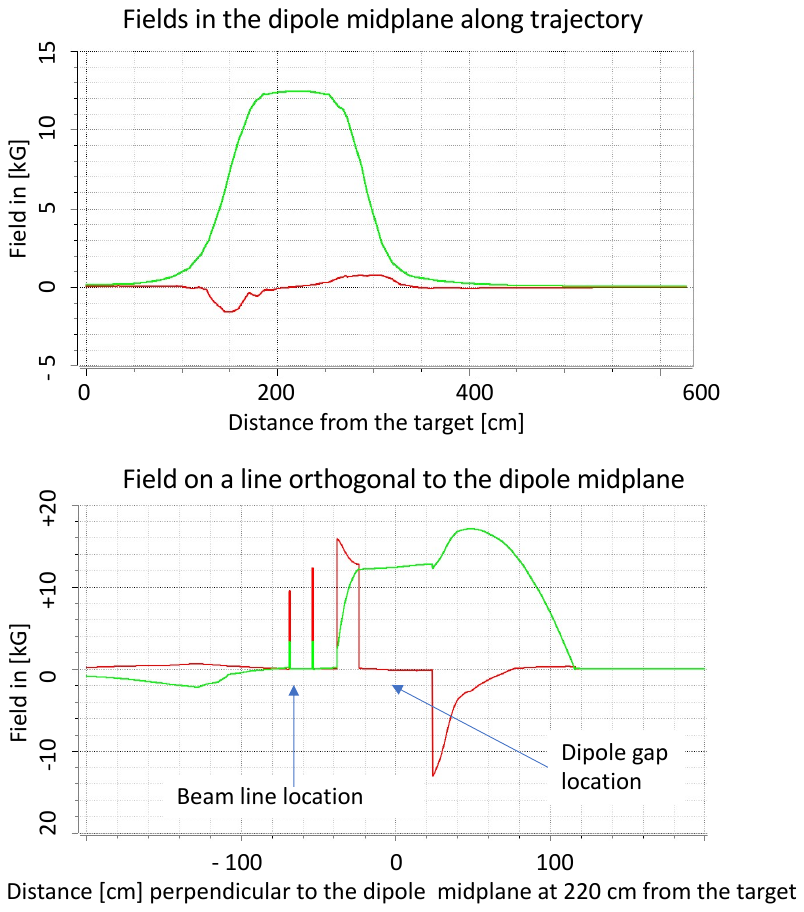}
\caption{The top panel shows the transverse (green line) and longitudinal (red line) field components relative to particle trajectory in the dipole midplane vs. distance from the target.
The bottom panel shows the field components (red line for the transverse and green line for the longitudinal) relative to the plane orthogonal to the dipole midplane at a distance of 220~cm from the target.}
\label{fig:DipoleField}
\end{figure}

The magnetic field (transverse and longitudinal relative to the particle trajectory in the dipole midplane) are shown in Fig.~\ref{fig:DipoleField} (upper panel).
The field in the gap (48~cm wide) is 12~kGauss while the field on the beamline is 1-2~Gauss near the middle of the magnet.
In an experiment which doesn't need full in-plane angular acceptance, additional pole shims could be added which would bring the field integral to 25~kGauss-meters.
Fig.~\ref{fig:DipoleField} (lower panel) shows the field components in the plane orthogonal to the dipole midplane,
which demonstrates effect of the magnetic shield of the beam line.

\subsection{\large Magnetic field along the beamline}
\label{sec:BeamLine}

\begin{figure}[htb]
\centering
\includegraphics[trim = 0mm 0mm 0mm 0mm, angle=0, width=0.8\textwidth]{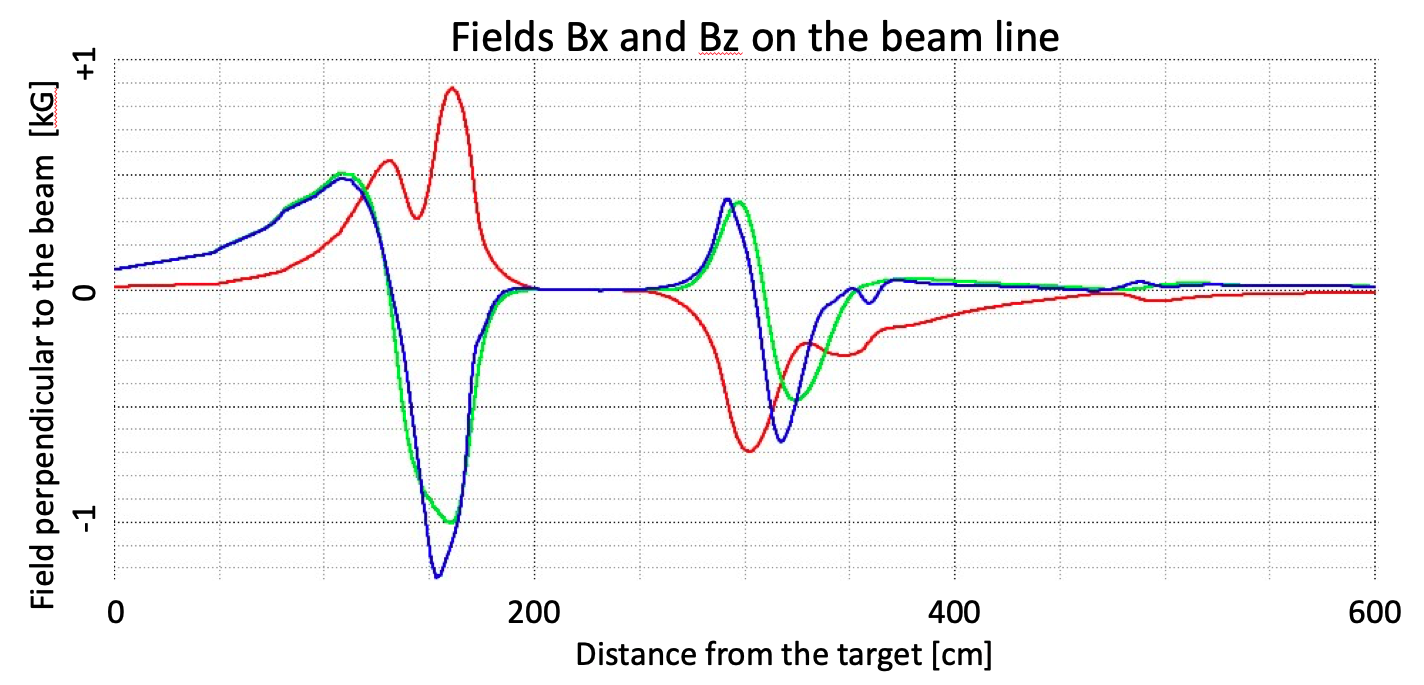}
\caption{Shown are the fields along the beamline for the central trajectory with both dipole correctors turned ON. 
The transverse (Bx) and longitudinal (Bz x 0.2) are shown as green and red lines respectively. 
Bx along a line tilted by 0.5 degree from the beam center, as a blue line.} 
\label{fig:BeamField}
\end{figure}

The beamline passes through the yoke in the horizontal cut which provides a very strong reduction of the field relative to the field in the gap.
The field components along the beamline are shown in Fig.~\ref{fig:BeamField}.
Here the $z$ axis is along the beam line; the $x$ and $y$ axes are in the horizontal and vertical directions, respectively.

Close to the ends of the yoke the field has a significant component transverse to the beam direction, \Bp, and additional shielding is necessary.
There are two requirements for the \Bp~integral, \BpI, along the beamline.
The first one is based on the acceptable beam deviation from the center of the beam dump aperture.
The \BpI~should be small enough that the beam deflection at the beam dump (about 40~m from the target) is less than 2~cm.
For a 10~GeV electron beam this leads to a limit on \BpI~of 20~kG-cm.
The second is based on radiation in the experimental hall and the counting rate in the detectors produced by secondary charged particles interacting with the beam pipe.
The SBS correctors allow us to reduce \BpI~below 3~kG-cm on the trajectories which have angles up to 0.5~degrees with the beam central line. 

\subsection{\large Beam line shielding}

\begin{figure}[!htb]
\includegraphics[trim = -100mm 0mm 0mm 0mm, angle=0, width=0.9\textwidth]{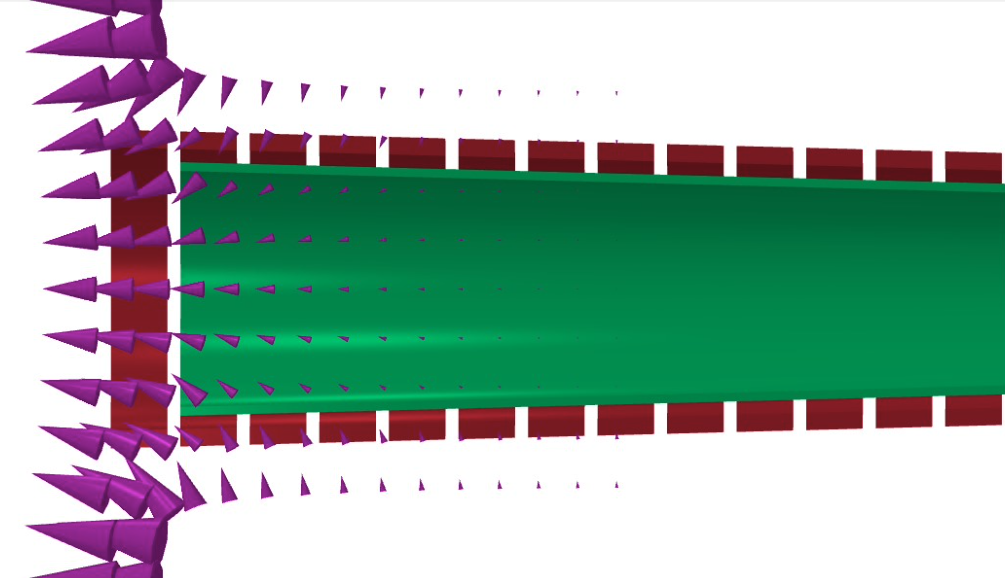}
\caption{The view of the vector field in the beamline shield near the exit from the SBS dipole (cut made for visibility of the structure).}
\label{fig:BeamLineShield}
\end{figure}

The field outside of the yoke is very strong and has both longitudinal and transverse components. 
The longitudinal component is not essential for the beam propagation but it saturates the shielding, which loses its ability to shield the transverse field.
The natural solution is based on solenoids for compensation of the longitudinal component, but during the experiment, the radiation level near the beam precludes such an option.
A sufficient shielding effect was found with passive shielding using a set of short tubes (rings) with small gaps between them, see Fig.~\ref{fig:BeamLineShield}.
Low-carbon steel AISI/SAE 1006 was used in all components of the beamline shielding.

\subsection{\large Beam line dipole correctors}

Field correctors were installed in front of and after the SBS dipole for minimization of the transverse field integral.
Fig.~\ref{fig:BeamField} shows the resulting field.
By using separated power supplies for the left and right coils of those correctors compensation was achieved in $\pm0.5^o$ angular range near the beam central line, which helps to reduce radiation induced by the low energy electrons emitted from the target. 

\subsection{\large Dipole field clamps}

A field clamp was used on the front of the magnet (see Fig.~\ref{fig:3D-view} and Fig.~\ref{fig:SBS-cut}) to reduce the fringe field in the target area to a level below 100 Gauss.  
A second field clamp was used in the rear of the magnet to minimize the field on the detector, especially on the tracking detector, which is located close to the dipole.

\subsection{\large Dipole gap shims}

\begin{figure}[!htb]
\includegraphics[trim = 30mm 30mm 80mm 30mm, angle=0, width=0.75\textwidth]{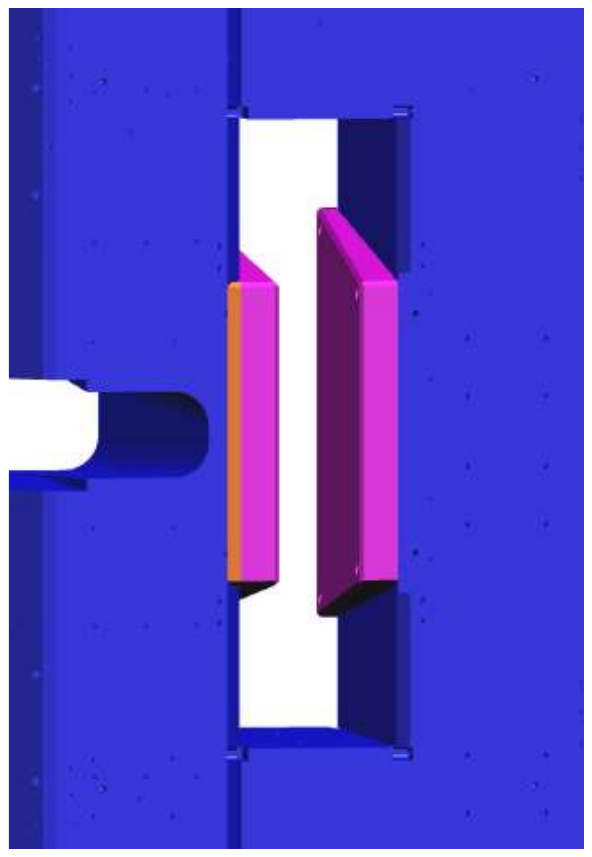}
\caption{The view of the yoke/dipole gap with the pole shims (shown in pink) used in the GEp experiment.}
\label{fig:Shims}
\end{figure}

Shims were used in the dipole gap (48~cm wide) for the GEp experiment, which required a solid angle of only 35 msr (out of a full 70~msr). 
The 3D view of the optimally shaped shim is shown in Fig.~\ref{fig:Shims}. 
With the shims the field integral in the dipole magnet increased by 25\%.

\subsection{\large Dipole coils}

The coil of the dipole on the side far from the beamline is a saddle type.
It has 140~turns and is designed for a maximum current of 2100~Amp.
The saddle coil was constructed by Buckley~\cite{NZ}.
The coils on the side which is close to the beamline have a flat shape to allow variation of the beamline position according to the desired angle of the spectrometer.

\subsection{\large The fringe field at the second detector arm}

In many experiments SBS was used together with a second arm detector located on the opposite side of the beamline. 
For example, in the GEp experiment the second arm was at 4~meters from the target with a central angle of 30~degrees.
At such a location the SBS fringe magnetic field is 100~Gauss oriented horizontally.

%% file: parameters.tex
\section{\large Spectrometer parameters}
\label{sec:parameters}

Many electron scattering experiments belong to one of two categories: 
i) one-arm, where only a scattered electron is detected, and \\ 
ii) two-arm, where both a scattered electron and a high momentum nucleon or meson are detected. 
At first glance, the second case seems to have much more demanding requirements for the spectrometers because acceptances of both arms need to be very large.
However, a particularly interesting physics case is when the recoiling hadron with high $z= p_h/p_{\gamma *}$ moves almost along the momentum of the virtual photon $\gamma*$ and has a small angle with respect to the incident electron beam.
In such a case, the required solid angle of the hadron arm is modest and the SBS spectrometer can provide it.
For example, for an electron beam of 11~GeV energy and the momentum transfer \qsq~of 12~\gevsq~at Bjorken variable $x_{_{BJ}} =1$, the recoiling proton momentum direction is at 16.9$^\circ$ with respect to the beam.
In a study of the semi-inclusive deep inelastic scattering at $x_{_{BJ}}\sim 0.4$, the central angle of the SBS would be 14$^\circ$~\cite{SIDIS}.
\begin{figure}[htb]
\includegraphics[trim = 50mm 60mm 50mm 30mm, angle=0, width=1.\textwidth]{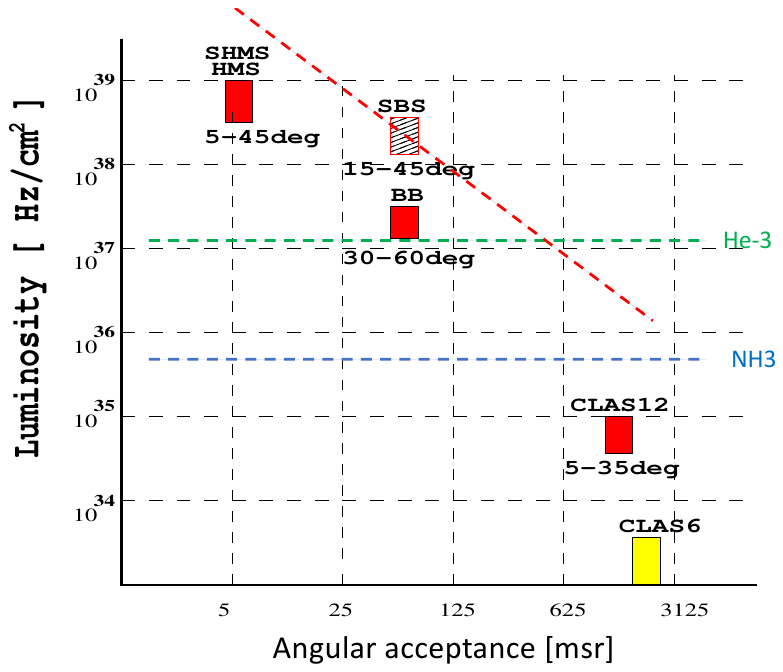}
\caption{Usable luminosity vs. solid angle of the JLab spectrometers. 
The green dashed line shows a limit for a polarized $^3$He~target. 
The blue dashed line shows a limit for a polarized NH3 target.
The red dashed line shows the achievable product of the luminosity and the detector solid angle in an experiment.}
\label{fig:LuminositySolidAngle}
\vskip -0.1 in
\end{figure}

The SBS spectrometer is a useful tool for a number of electron scattering experiments when the solid angle is important.
A large group of such experiments uses polarized targets which can operate at electron-nucleon luminosity above $10^{34}$~cm${^{-2}}$/s.

In Fig.~\ref{fig:LuminositySolidAngle} we show the usable luminosity as a function of angular acceptance for several JLab spectrometers.  
The dashed red line corresponds to a constant value of the product of luminosity and solid angle and illustrates the superior performance of SBS for an important class of experiments.  
Also shown in Fig.~\ref{fig:LuminositySolidAngle} are green and blue horizontal lines indicating the maximum luminosities at which polarized He-3 and polarized NH3 targets can operate, respectively.

There are five main parameters characterizing the spectrometer: the momentum and angular resolutions, the momentum and angular acceptances, and the minimal central angle with respect to the beam.

The relative momentum resolution of the SBS spectrometer is (0.3 + 0.03$\times$p[GeV])\%. It is mainly defined by the field integral in the dipole and the combination of the angular resolution by the tracking detector and multiple scattering of the particle in the material in its path. 
The momentum acceptance of SBS is large because the deflection angle is small compared with the vertical angular acceptance.
The minimum central angle with respect to the beam is limited by the selected magnetic elements due to its size and fringe field on the beamline.
Table~\ref{tab:SBS_parameters} shows the solid angle of the SBS spectrometer at different central angles.

It is interesting to note that for the angles below 15 degrees the SBS solid angle covers a significant portion ($\sim$25\%) of the maximum possible solid angle, defind as 
$2\pi\times \sin(\theta_{hor})\times \Delta \theta_{hor}$.
\begin{table}[ht]
\begin{center}
\vskip -0.2 in
\renewcommand{\arraystretch}{1.05}
\caption{Solid angle of the SBS spectrometer vs. the central angle.}
\vskip 0.1 in
\label{tab:SBS_parameters}
    \begin{tabular}{|c|c|c|c|c|c|}
\hline  
$\theta_{central}$ & $\Omega$ & D & Hor. range & Vert. range & $\Omega$ fraction\\ 
degree  &  msr & meter & degree & degree & of $2\pi \cdot \theta \Delta \theta$\\ 
\hline
3.5 & 5. & 9.5 & $\pm$ 1.3 & $\pm$ 3.3  & 0.28 \\  
5.0 & 12 & 5.8 & $\pm$ 1.9 & $\pm$ 4.9  & 0.33 \\  
7.5 & 30 & 3.2 & $\pm$ 3.0 & $\pm$ 8.0  & 0.35 \\
15. & 72 & 1.6 & $\pm$ 4.8 & $\pm$ 12.2 & 0.26 \\ 
30. & 76 & 1.5 & $\pm$ 4.9 & $\pm$ 12.5 & 0.14 \\ 
\hline  
    \end{tabular}
\end{center}
\end{table} 

%% file: detectors.tex
\section{\large Detectors}
\label{sec:Detectors}

The Super Bigbite Spectrometer allows the use of a large and flexible detector package. 
The list of components in the first group of experiments~\cite{GEp, GMn, GEn, nTPE, GEn-RP}
includes a front tracking detector, a rear tracking detector, and a hadron calorimeter. 
In some future experiments~\cite{SIDIS, TDIS} we also plan to use a highly segmented Cherenkov counter (RICH)~\cite{RICH} and a lead-glass shower calorimeter~\cite{HERMES} (in combination with a hadron calorimeter for particle identification), both reused from HERMES.
Here we present two detectors constructed specifically for the GEp experiment~\cite{GEp, GEp-update}.

\begin{figure}[htb]
\centering
\includegraphics[trim = 40mm 40mm 0mm 30mm, angle=0, width=1.2\textwidth]{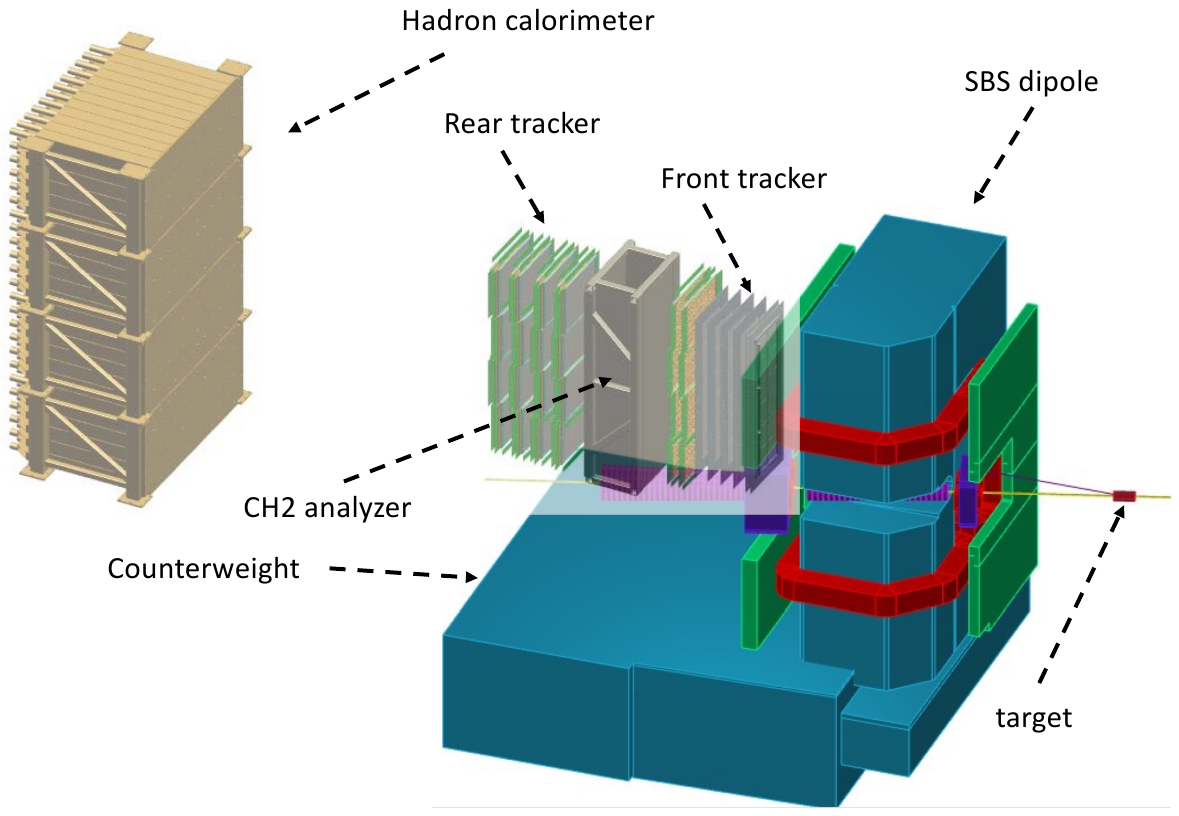}
\caption{The SBS spectrometer as used in the GEp experiment.} 
\label{fig:SBS-GEp}
\end{figure}

The layout of the GEp detector package is shown in Fig.~\ref{fig:SBS-GEp}.

Detectors provide information on the particle trajectory, time of arrival and its energy.
In the GEp experiment we used two tracking detectors, in front of and behind the CH2 analyzer, which allowed us to find the scattering angle in the CH2 analyzer needed for measurement of the proton polarization~\cite{Basiliev,Azhgirey}.

\subsection{\large Tracking detector}

GEM-based coordinate detectors of similar design have been constructed for the front and rear tracking detectors with a 56-cm CH2 block between them.
Each tracking detector includes 8~planes of GEM-based chambers~\cite{GEM}.
The detector dimensions vary: 40~cm x 150~cm in six front planes and 60~cm x 200~cm in the remaining 10 planes; these 10 larger layers are made up by combining four 60 cm x 50 cm GEM modules into each layer. 

Readout is arranged with 50-60~cm long strips of 0.4~mm pitch.
Each chamber provides coordinate resolution of 70~$\mu$m.
Orientation of the strips varies (90, 45, 30, 0 degrees) to help with the straight track search.

\subsection{\large Calorimeter}

The hadron calorimeter~\cite{HCAL} is located behind the tracking detectors at a distance of 10-17~m (in different experiments) from the target.
The large area of the calorimeter of 1.8~m by 3.6~m allows us to use it as a trigger in the case of the GEp experiment.
The calorimeter consists of 288 modules of alternating steel-scintillator layers. 
Each module has dimensions 15.5~cm x 15.5~cm x 100~cm with a fast wavelength shifter plate in the middle.
The fast scintillator with 0.5\% PPO (2,5-diphenyloxazole) was fabricated at Fermi National Accelerator Laboratory by the extrusion method.
Photons emitted from the fast scintillator in a module were absorbed in a fast wavelength shifter which ran the 100~cm length of the module and bisected the 15.5~cm x 15.5~cm planes.   
The resulting wavelength-shifted photons were guided to the downstream end of the module and the signal was read out with a single PMT on each module.
The light collection to the PMT used a novel shape adiabatic light guide~\cite{ALG}.
XP2262 and XP2282 PMTs are used for light detection.
Time resolution for high energy proton detection in this calorimeter was found from experimental data to be 0.75~ns (rms).